\documentclass{ws-procs9x6}
\usepackage{epsfig,graphicx,wrapfig}
\def\lsim{\mathrel{\rlap{\lower4pt\hbox{\hskip1pt$\sim$}}
    \raise1pt\hbox{$<$}}}         

\begin{document}
\pagestyle{empty}
{\large
\begin{center}

{\Large\bf SPIN PHYSICS}\\
\bigskip
\bigskip
{\large\bf  Stefano Forte}\\
\smallskip
{ \it Dipartimento di Fisica, Universit\`a di Milano and\\ 
INFN, Sezione di Milano,\\ Via Celoria 16, I-20133 Milano, Italy}\\
\bigskip
{\large\bf  Yuji Goto}\\
\smallskip
{ \it RIKEN,\\
Wako, Saitama 351-0198, Japan and\\
RIKEN-BNL Research Center\\
Upton, New York 11973-5000, U.S.A.}

\vskip1.cm

{\bf Abstract}\\
\end{center}
We review recent experimental and theoretical progress in
  spin physics, as presented in the spin parallel session of
  DIS2006. In particular, we discuss the status of the nucleon spin
  structure, transverse polarized asymmetries, and recent developments
  such as DVCS, polarized fragmentation and polarized resummation.

\noindent

\vskip1.5cm
\begin{center}
Convenors' Summary  Talk at \\
{\bf DIS 2006}\\
Tsukuba, Japan, April  2006\\
{\it to be published in the proceedings}
\end{center}
\vfill
IFUM-868/FT\\
RIKEN-AF-NP-472\hfill  July 2006} 
\eject
\setcounter{page}{1} \pagestyle{plain}

\title{Spin Physics}

\author{Stefano Forte}

\address{
Dipartimento di  Fisica, Universit\`a di
Milano and\\INFN, Sezione di Milano, Via Celoria 16, I-20133 Milan, Italy}

\author{Yuji Goto}

\address{RIKEN, 
Wako, Saitama 351-0198, Japan\\
RIKEN BNL Research Center,
Upton, New York 11973 - 5000, U.S.A.}  

\maketitle

\abstracts{We review recent experimental and theoretical progress in
  spin physics, as presented in the spin parallel session of
  DIS2006. In particular, we discuss the status of the nucleon spin
  structure, transverse polarized asymmetries, and recent developments
  such as DVCS, polarized fragmentation and polarized resummation.}

\section{The polarized structure of the nucleon}
\label{polstr}

Experimental and theoretical studies of spin physics in the last
several years have considerably widened their scope. 
Inclusive polarized deep-inelastic measurements, and their
interpretation in terms of polarized quark and gluon structure
functions, are now supplemented by measurements of semi--inclusive
processes, heavy quark production and high--$P_T$ hadron production
and deeply--virtual Compton scattering (DVCS) 
in lepton--nucleon scattering, by data collected in a variety of
hard processes at the polarized hadron collider RHIC, and by data on
polarized fragmentation from $e^+e^-$ machines. Their 
interpretation requires both a deepening and a widening of available
theoretical tools. The wealth of new data on the
spin structure of the nucleon requires the use of the more advanced
techniques that are being developed in the unpolarized case for the
description of the parton structure of hadrons, specifically in view of
the LHC:\cite{Dittmar:2005ed} higher order QCD computations, resummation,
global parton fits with errors. Also, new quantities must
be introduced, along with their theoretical interpretation within QCD:
polarized fragmentation functions, 
transverse momentum distributions and orbital angular momentum,
transverse spin distributions and their cognates.

In this brief review, based on the presentations in the spin working
group at DIS06, we will first review the status of the nucleon spin
problem: we will summarize new determinations of the polarized parton
distributions $\Delta q$ and $\Delta g$ in lepton scattering at
CERN and DESY and in proton--proton scattering at RHIC, and first data
on DVCS from HERMES, and we will discuss their theoretical analysis and
interpretation. We will then summarize recent progress on transverse
spin asymmetries: we will review several recent
asymmetry measurements in hadron production at CERN, DESY and RHIC,
and we will review recent progress in the
formulation of a unified approach to transverse single--spin
asymmetries based on perturbative
factorization. Finally, we will discuss several recent new
developments which extend the range of experimentally accessible
quantities and computational techniques to the polarized
case: specifically, we will analyze measurements of polarized
fragmentation (BELLE and COMPASS) and  structure functions at low $Q^2$
(JLab experiments), and discuss the
development of polarized resummation methods.

\section{The nucleon spin puzzle}
\label{spinpuz}

As well known, the nucleon spin problem\cite{ridolfi} has to do with
the  fact that, loosely speaking, the measured quark spin
fraction is small. One may wonder why
this is a problem: given that the nucleon mass is not
carried by the quark masses, and only about half of it is due to
quark interactions, why should the nucleon spin be carried by the quark
spin? 
The answer is, of course,\cite{Veneziano:1989ei} that what is
surprising is  the
violation of the OZI rule: nucleon matrix elements of the singlet axial
current are much smaller than those of the octet, i.e.
$a_0=a_u+a_d+a_s<\!\!<a_8=a_u+a_d -2 a_s$,
where the axial charges $a_i$ are just the forward quark current matrix
elements from flavor $i$: $\langle N;\, p,s|{
J^\mu_{5,\, i}}|N;\, p,s\rangle={ a_i} M_N {
s^\mu}$. 

Explanations of this fall in two broad classes: either
the singlet is special, because  it can
couple to gluons, or the octet is special,
because strangeness in the nucleon is much larger than one might
expect. Hence, in order to understand the spin puzzle one needs
precise measurements of the gluon, strange and antistrange quark
distributions, specifically their first moments. 

Also,
$a_0$ and $a_8$ currently are not  determined directly, but rather
from the combination of a direct measurement and the indirect determination of an independent
linear combination, obtained using SU(3)  from baryon octet 
beta--decay rates. A direct measurement of the nucleon
axial charge for each flavor would be desirable: this, in turn,
entails the determination of the first moment of all polarized light
quark and antiquark densities, as well as of the polarized gluon
density, that mixes in the singlet.

\subsection{Experimental results on $\Delta q$ and $\Delta G$}
\label{expdqdg}
New experimental results relevant for the determination of polarized
parton distributions have been obtained recently both in lepton DIS
(COMPASS and HERMES) and in proton--proton scattering (STAR and PHENIX).

At the inclusive DIS level, the 
COMPASS experiment presented $A_1^d$ and $g_1^d$ results for $Q^2 > 1$
GeV$^2$.\cite{stolarski}
This new results improves their QCD analysis, and it gives
$\Delta \Sigma = 0.25 \pm 0.02$ (stat) and $\Delta G = 0.4 \pm 0.2$ (stat)
at $Q^2 = 3$ GeV$^2$.

At the semi-inclusive level, the 
HERMES experiment obtained updated $\Delta s + \Delta \bar{s}$
distribution from their DIS measurement and semi-inclusive $K^+ + K^-$
measurement with the polarized-deuterium target.\cite{ehrenfried}
Since the strange quarks carry no isospin and the deuteron is an isoscalar
target, they obtained $\Delta s + \Delta \bar{s}$ with two assumptions,
isospin symmetry between proton and neutron, and charge-conjugation
invariance in fragmentation.
They also obtained the fragmentation functions needed in this analysis from
multiplicities directly at HERMES kinematics with the same data.
The result shows that for  the $K^+ + K^-$ fragmentation function from
non-strange quarks
 the strangeness suppression factor for $s + \bar{s}$
production is important.
The $\Delta s + \Delta \bar{s}$ distribution is consistent
with zero with improved uncertainties.
\begin{figure}[ctb] 
\begin{center}
\includegraphics[width=.55\linewidth]{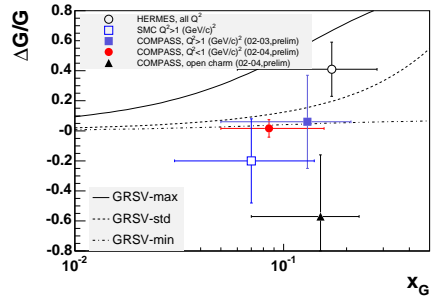}
\end{center}
\vspace{-.2cm}
\caption{The gluon polarization obtained by COMPASS from
high-$p_T$ hadron pairs and open charm production,
compared to  SMC and HERMES determinations 
from  high-$p_T$ hadron pairs.}
\label{fig:DeltaG-COMPASS}
\vspace{-.4cm}
\end{figure}

COMPASS can access  $\Delta G$ directly by
three methods, high-$p_T$ hadron-pair measurement at low $Q^2$ and at high
$Q^2 > 1$ GeV$^2$, and open charm production.\cite{kurek}
Results are summarized  in 
Fig.\ref{fig:DeltaG-COMPASS}: either $\Delta G$ is small or it has a
node
at $x \sim 0.1$. However, a first moment of
$\Delta G \sim 0.4$ at low scale  is not excluded.

\begin{figure}[ctb] 
\begin{center}
\includegraphics[width=.42\linewidth]{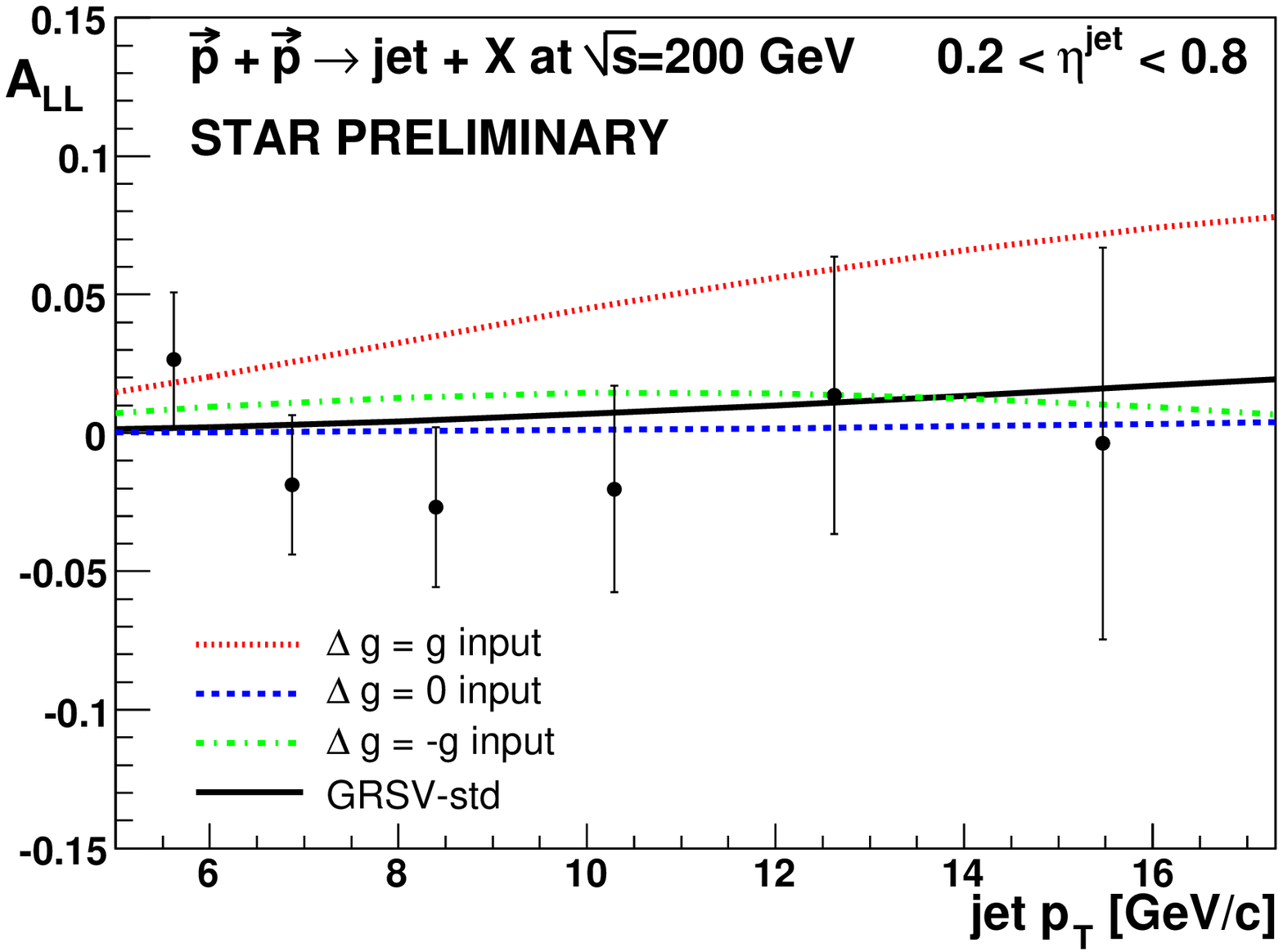}
\includegraphics[width=.42\linewidth]{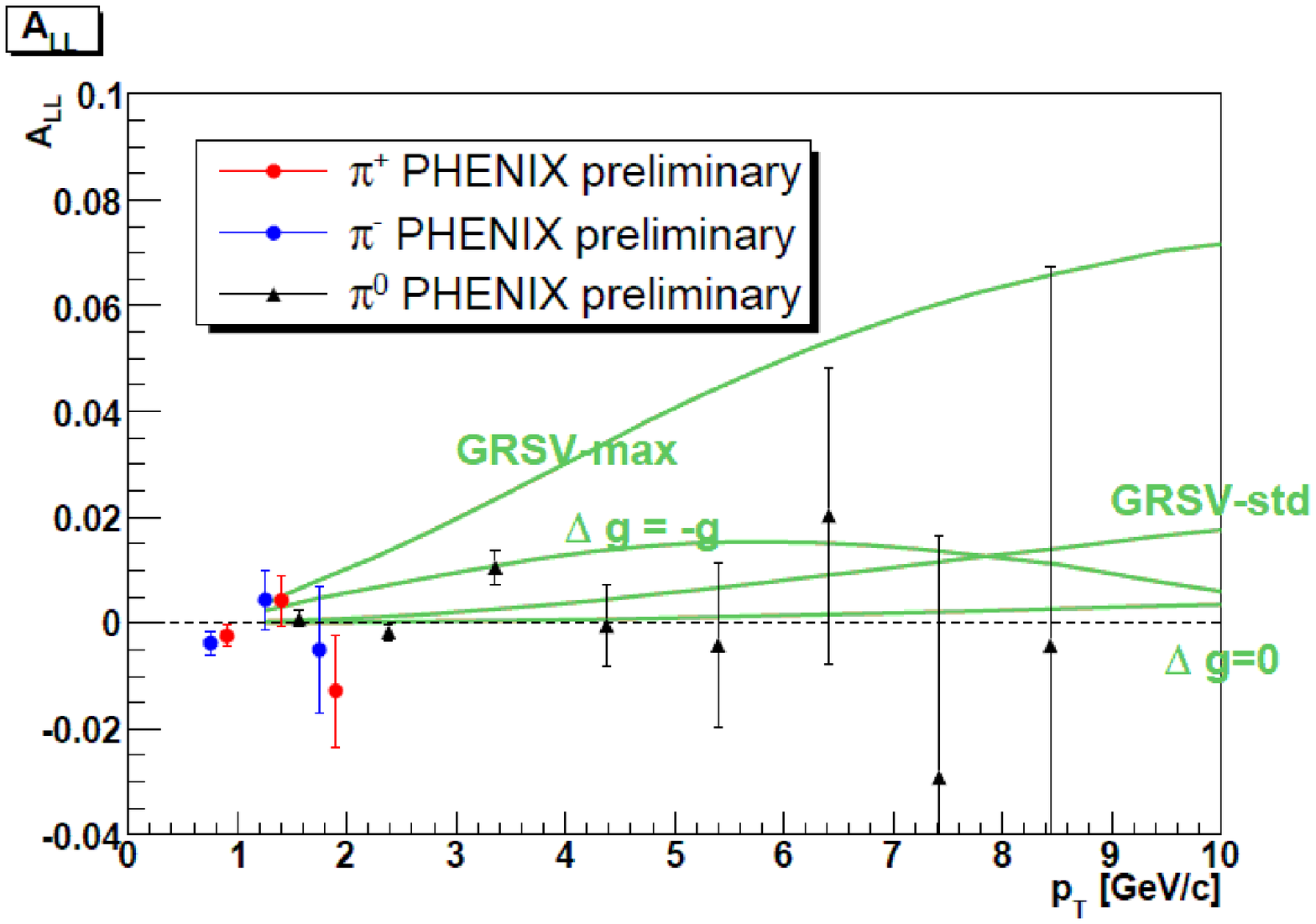}
\end{center}
\caption{The double helicity asymmetry in the inclusive jet production
measured by STAR (left) and that in the inclusive $\pi^0$
production measured by PHENIX (right) at midrapidity in
polarized $p + p$ collisions.}
\label{fig:DeltaG-RHIC}
\vglue-.6cm
\end{figure}

First results relevant for the determination of $\Delta G$ were
recently obtained by RHIC experiments.
The STAR experiment presented their preliminary result of the double
helicity asymmetry in inclusive jet production at midrapidity in
polarized $p + p$ collisions.\cite{kiryluk}
The result shown in Fig.\ref{fig:DeltaG-RHIC} is limited by
statistical uncertainties and it cannot distinguish between different scenarios
for the gluon polarization, though it 
disfavors a large value of $\Delta G$. 
The PHENIX experiment presented results for the 
double helicity asymmetry of 
inclusive $\pi^0$ production at midrapidity in polarized $p + p$
collisions.\cite{fukao}
The result, shown in Fig.\ref{fig:DeltaG-RHIC},
also disfavors  large $\Delta G$ and excludes scenarios where $\Delta
G$ is as large as $G$ at low scale (GRSV-max) .


\subsection{Orbital angular momentum and DVCS}
\label{dvcs}

\begin{figure}[ctb]\vglue-.6cm 
\begin{center}
\includegraphics[width=.5\linewidth]{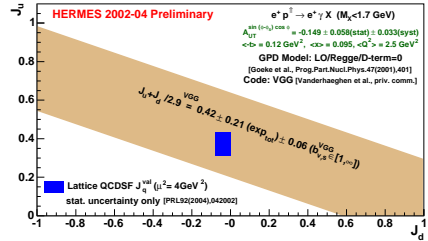}
\end{center}
\caption{A model-dependent constraint on $J_u$ and $J_d$ obtained by
HERMES and comparison with a quenched lattice-QCD calculation.}
\label{fig:DVCS}
\vglue-.5cm
\end{figure}

The only known way to access  experimentally the 
orbital angular momentum is through a measurement of  generalized
parton distributions (GPDs).
GPDs can be determined e.g. in deeply virtual Compton
scattering (DVCS).
HERMES has provided a first determination of
 the transverse target-spin
asymmetry associated with DVCS, $A_{UT}(\phi,\phi_S)$,
on the proton.\cite{ye}
The $\sin(\phi-\phi_S) \cos(\phi)$ term of $A_{UT}(\phi,\phi_S)$ is sensitive
to $J_q$.
A model-dependent constraint on $J_u$ and $J_d$ was obtained by comparing
the asymmetry and the theoretical predictions based on a GPD model.
Figure \ref{fig:DVCS} shows the result and comparison with a quenched
lattice-QCD calculation.

\subsection{The state of the art: partial results and global fits}
\label{spinpuzth}

\begin{figure}[b]
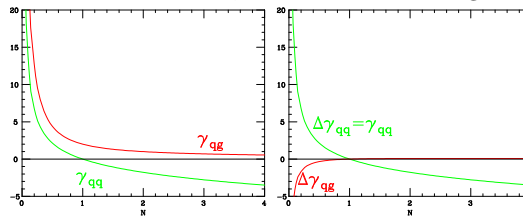
 \vglue-.6cm \begin{center}
\includegraphics[width=.3\linewidth]{gs.ps}
\includegraphics[width=.3\linewidth]{dgs.ps}\end{center}
\caption{Unpolarized (left) and polarized (right) anomalous dimensions.
\label{fig:ad}}
\vglue-.6cm
\end{figure}
Inclusive deep-inelastic experiments can only lead to the
determination of one linear combination of quark plus antiquark
polarized densities, $g_1\sim\sum_i e^2_i\left(q_i+\bar q_i\right)$ ---
two, if proton and deuteron targets are available. Also, they
only provide a weak handle on the gluon through scaling
violations, due to the smallness of the relevant first moments of
polarized anomalous
dimension (see Fig~\ref{fig:ad}). Hence, they only determine
accurately the isotriplet (Bjorken sum rule), while the
singlet quark and gluon first moments are affected by large
uncertainties, and there is essentially  no information on total 
strangeness.\cite{ridolfi} Recent more precise inclusive data\cite{stolarski}
further improve the Bjorken sum rule and provide some more information
on the small $x$ behavior of the $g_1$ structure function, but cannot
change this situation.

This is why recent effort has concentrated on 
less inclusive observables. Semi--inclusive
deep--inelastic scattering (SIDIS) seems especially useful for the determination
of  individual polarized flavors and antiflavors. At leading
perturbative order, one can form combinations of measurable
asymmetries which are independent of fragmentation  and thus measure
polarized flavors and antiflavors directly: specifically, the strange
polarized distribution.\cite{ehrenfried} 
\begin{figure}[ctb]  \begin{center}
\includegraphics[width=.7\linewidth]{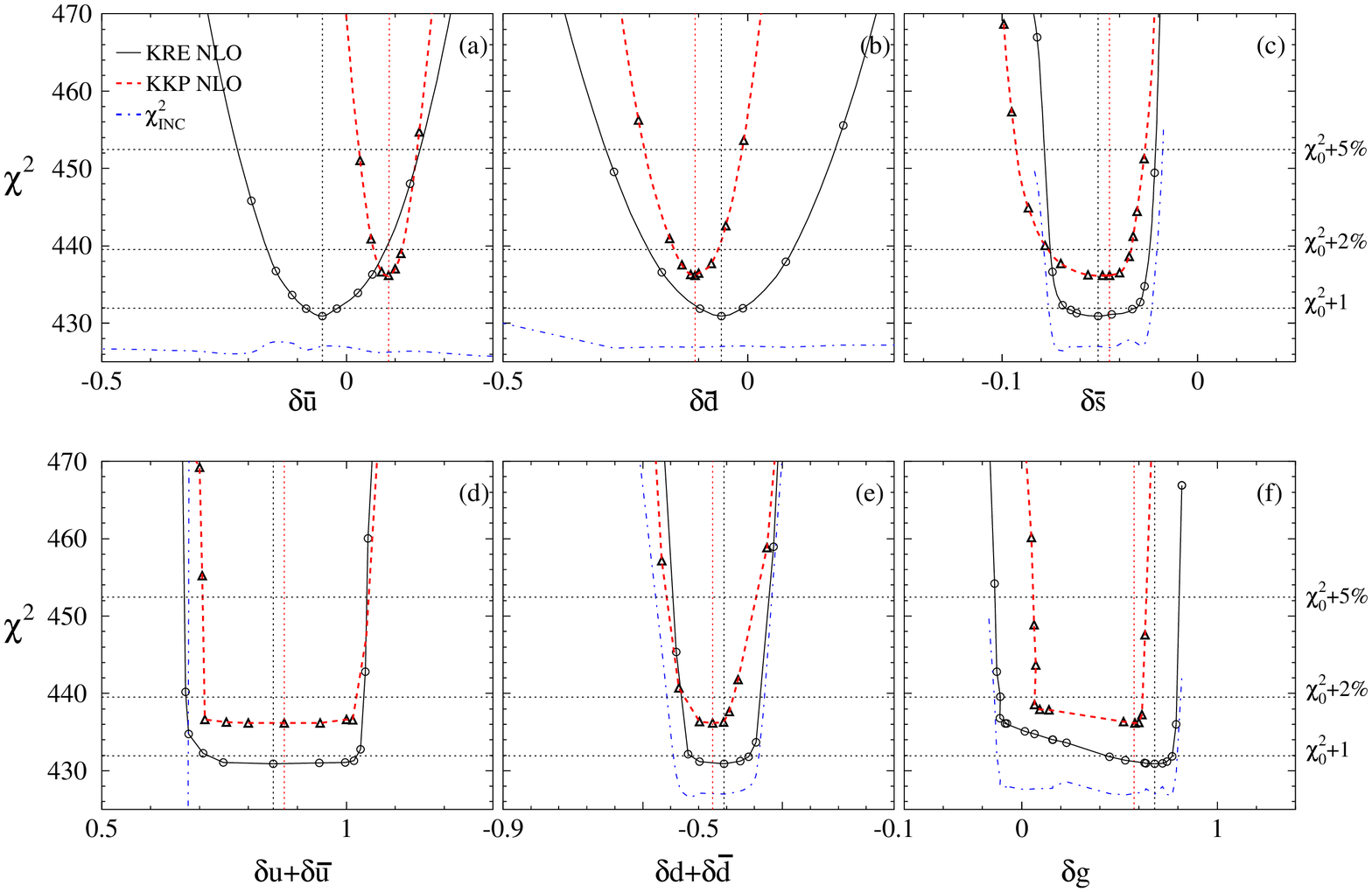}\end{center}
\caption{Global fit to inclusive and semi--inclusive deep--inelastic
  scattering data .
(From Ref.~\protect\cite{sassot})
\label{fig:dissidis}}
\vglue-.5cm
\end{figure}

However, 
such a
leading--order analysis is only accurate if the dominant contribution
to fragmentation into a given hadron comes from the quark carrying the
corresponding flavor quantum number: i.e., it assumes the validity of
the very OZI rule whose violation  we are trying to understand.
Indeed, a full  NLO fit including all available DIS and SIDIS data\cite{sassot}
(see Fig~\ref{fig:dissidis})
shows that e.g. the first moment of the anti-up distributions changes
by a factor two from LO to NLO, and can even change sign according to
the choice of fragmentation functions. 
\begin{figure}[b]
 \begin{center}
\includegraphics[width=.57\linewidth]{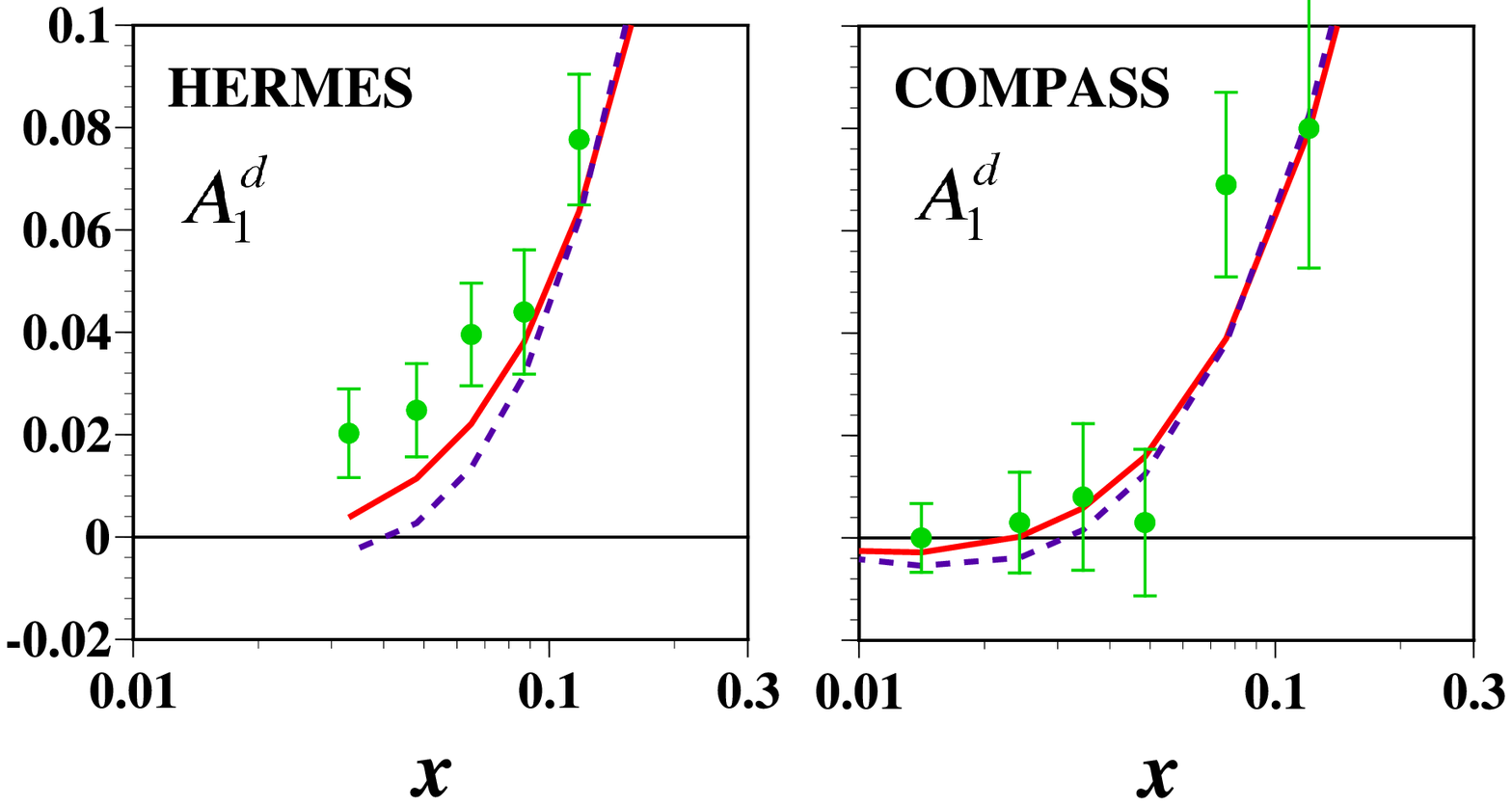}
\includegraphics[width=.37\linewidth]{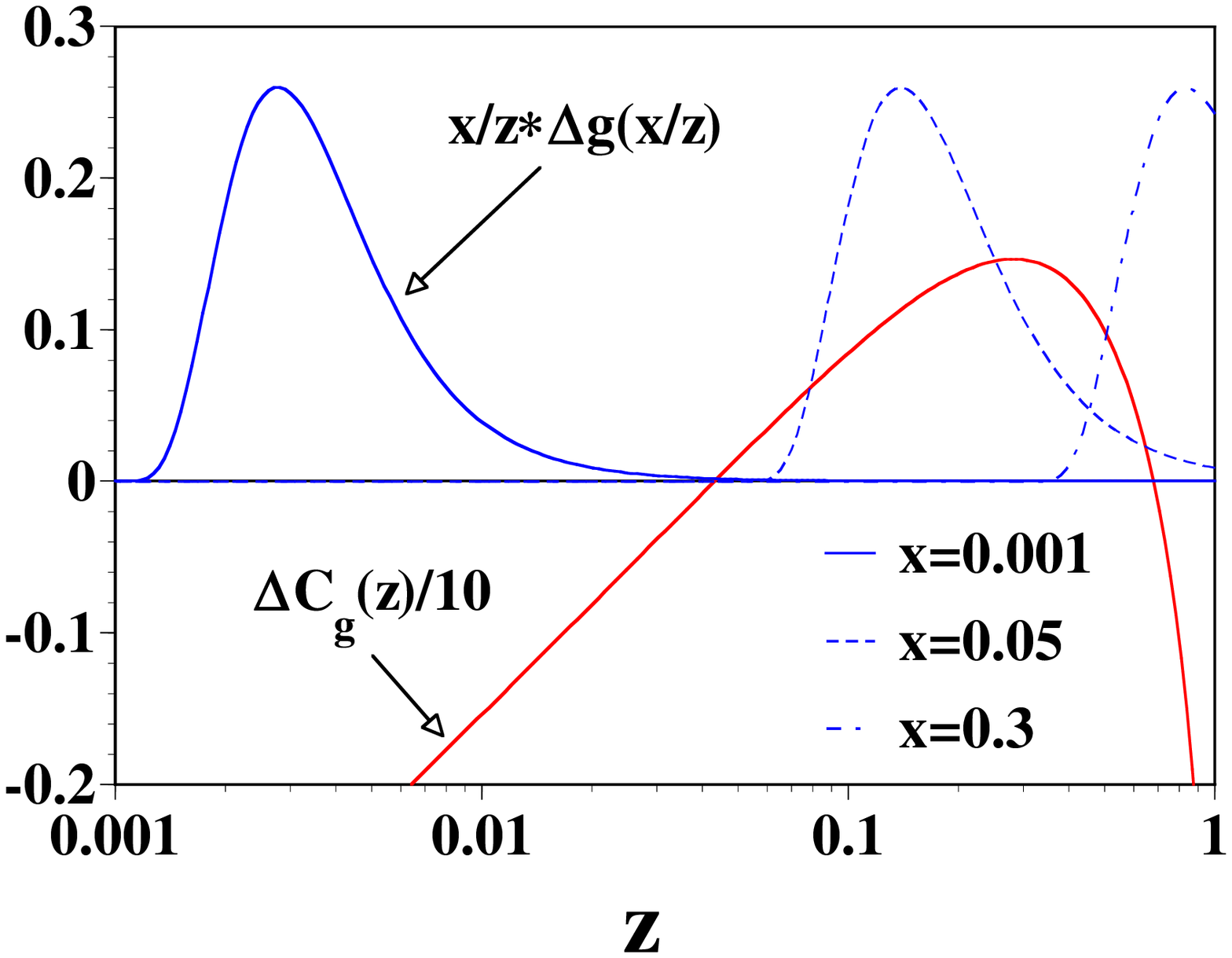}\end{center}
\caption{Possible evidence for large $\Delta g(x)/g(x$ at large $x$
  from HERMES and COMPASS data. (From Ref.~\protect\cite{hirai})
\label{fig:aacdf}}
\vglue-.7cm
\end{figure}

\begin{wrapfigure}{r}{.4\linewidth}
\vglue-.3cm
\includegraphics[width=\linewidth]{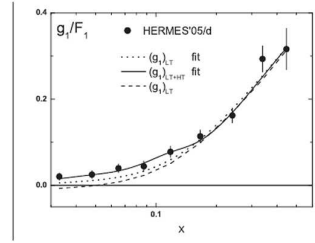}
\vglue-.15cm
\caption{Possible evidence for higher twist contributions from HERMES data. 
(From Ref.~\protect\cite{leader})
\label{fig:leadht}}\vglue-.6cm
\end{wrapfigure}

Analogously,  there are intriguing suggestions that available
information may provide a handle on $\Delta g$: for example, a
shortfall of a leading--order determination of inclusive DIS polarized asymmetries 
at medium--small $x\sim 0.05$ may be related\cite{hirai} to the presence of a sizable
positive $\Delta g(x)$  at large $x\sim 0.3$ (see
Fig.~\ref{fig:aacdf}). However, the same effect can also be explained
by higher twist contributions\cite{leader} (see Fig.~\ref{fig:leadht}).

\begin{figure}[b] \vglue-1.cm
 \begin{center}
\includegraphics[width=.42\linewidth]{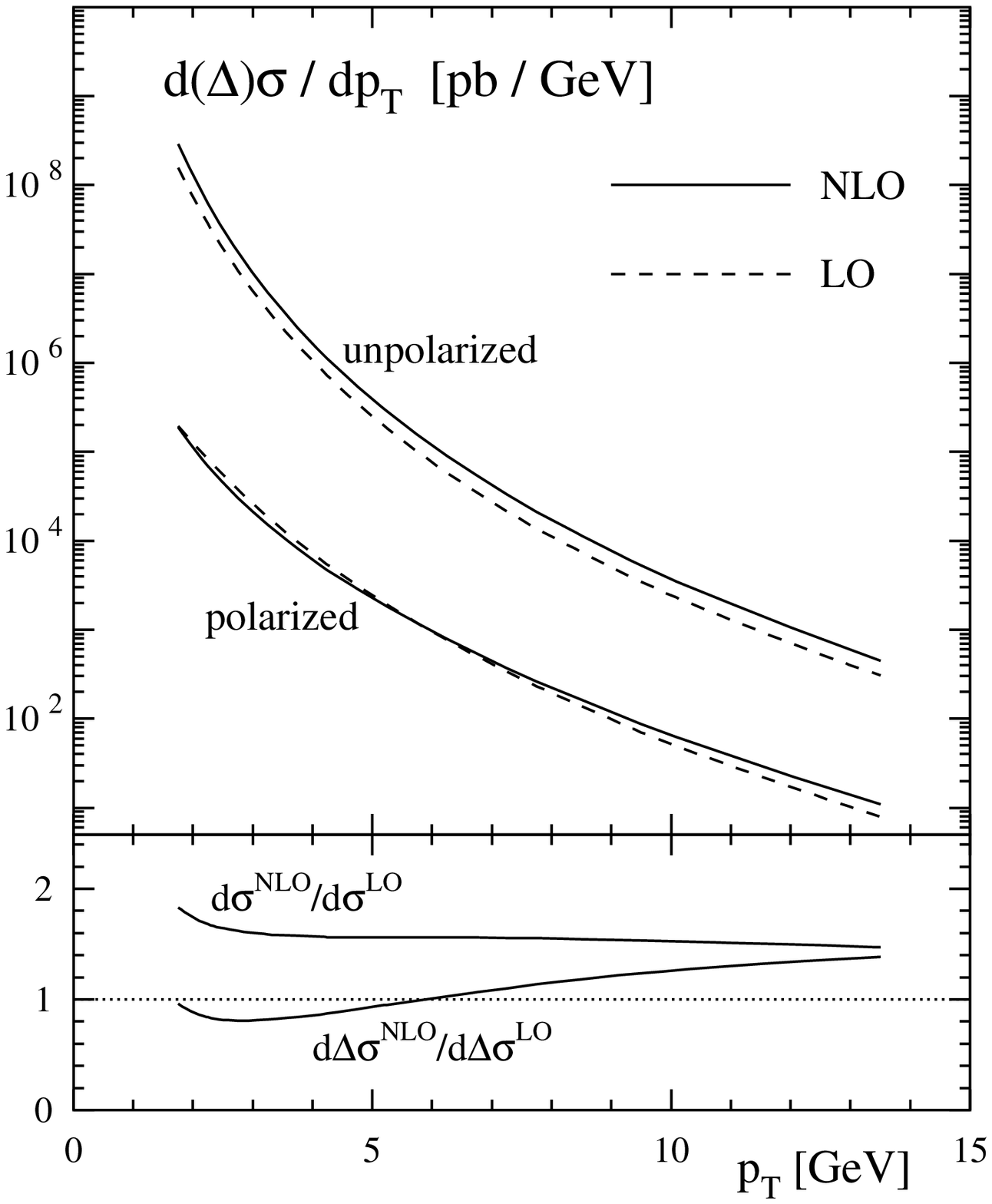}
\includegraphics[width=.42\linewidth]{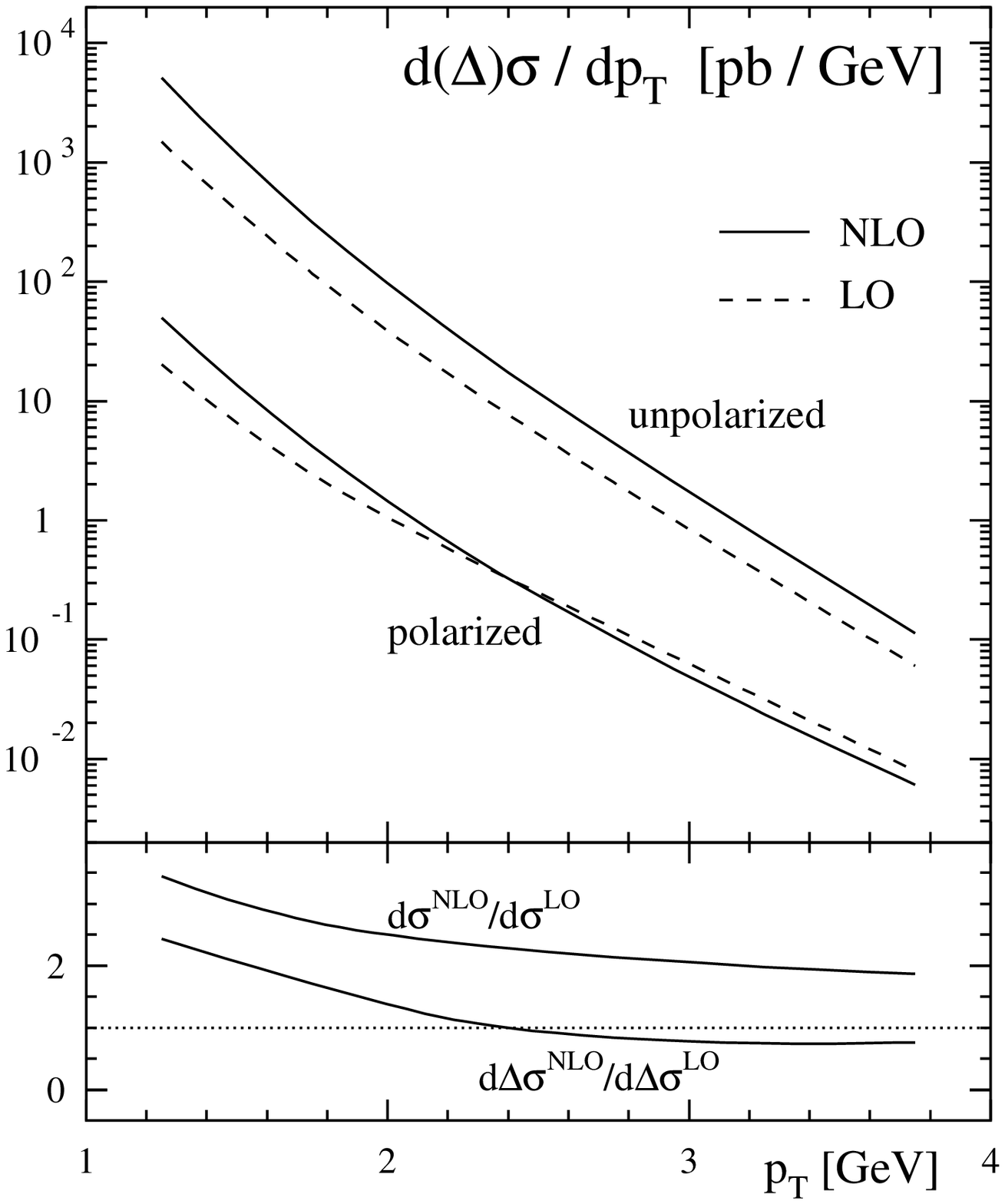}\end{center}
\caption{Cross--sections and $K$--factors for single--inclusive pion
  production at RHIC (left) at COMPASS (right).
(From Ref.~\protect\cite{stratmann})
\label{fig:kfact}}
\vglue-.5cm
\end{figure}
Whereas RHIC provides us with many processes which are sensitive to
$\Delta g$, and for which higher twist corrections are small and no
further non--perturbative input (such as from fragmentation functions)
is required, it remains true that NLO corrections are generally quite
large. In particular, polarized $K$--factors are neither
small nor constant (see Fig.~\ref{fig:kfact}), and therefore the full
available NLO information must be used.\cite{stratmann}

\begin{wrapfigure}{r}{.4\linewidth}   
\vglue-.5cm
\includegraphics[width=.6\linewidth,angle=90]{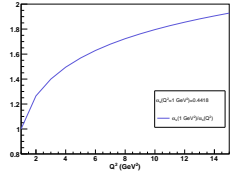}
\caption{Leading--order scale dependence of the first moment of
  $\Delta G$.\label{fig:dgscal}}
\vglue-.8cm
\end{wrapfigure}

Furthermore, it is important to remember that 
the scale dependence of the first moment of $\Delta G$  is quite
  strong: at leading order, it   evolves as
  $\frac{1}{\alpha_s(Q^2)}$, so that e.g. it varies by a factor two
  when the scale is increased from, say $1$ to $15$~GeV$^2$ (see
  Fig.~\ref{fig:dgscal}). This, together with the potentially large
  $K$--factors,  should be kept in mind when comparing different
  determinations of $\Delta G$, such as in fig.~\ref{fig:DeltaG-COMPASS}
Therefore, 
  the interesting COMPASS charm production data
are likely to  play a significant role in the determination of $\Delta
G$ only if the corresponding NLO corrections (presently available for
polarized charm photoproduction and hadroproduction, but not
electroproduction) will become available, while double--inclusive
  large $p_T$ hadrons are unlikely to play an important role in the
  near future.

Because the very many independent processes which will soon be
measured at RHIC are sensitive to both the gluon and individual light
flavors and antiflavors, 
it is likely
that both a direct determination of $a_0$, $a_8$ and a handle on the first moment
of $\Delta g$ will soon be possible. Yet,  the separation of polarized
strangeness and antistrangeness will probably only be possible at a
neutrino factory.
and a precise determination of $\Delta G$ including the small $x$
region 
will have
to wait for eRHIC.
However, it will only be possible to extract  this information from
the data through a global fit, and no single experiment is likely to be
sufficient.

The conclusion on $\Delta s$ and $\Delta G$ from present--day data is
therefore not very different from what it was after the latest and
most precise inclusive DIS data:\cite{ridolfi} the first moment of the
polarized gluon is likely to be positive, though there are some
indications that it might be smallish (i.e. $\left(\frac{n_f}{2\pi}\right)\alpha_s(Q^2)\Delta G
\lsim\frac{1}{2}a_0$), and the
first moment of the total strangeness is likely to be negative.

\section{Transverse spin asymmetries}
\label{transv}
Many new determinations of transverse spin asymmetries have been
performed recently, both in lepton proton and proton--proton
scattering. Correspondingly, significant progress has been made in the
interpretation of these measurements.

\subsection{Experimental results}
\label{transvexp}

Collins moments and Sivers moments of the
transverse single-spin asymmetry (SSA) with a polarized-proton target for
semi-inclusive charged pions and kaons have been determined by
HERMES.\cite{pappalardo}
The Collins moments of $\pi^+$  and $\pi^-$ 
have positive and negative asymmetries respectively, that of $K^+$ has a small
or zero asymmetry, and that of $K^-$ has a positive asymmetry.
The Sivers moments of $\pi^+$ and $K^+$ 
have positive asymmetries, and those of $\pi^-$ and $K^-$ show small or zero
asymmetries.
These results support the existence of non-zero chiral-odd and T-odd
structures that describe the transverse structures of the nucleon.
First measurement for kaons suggest that sea quarks may provide an important
contribution to the Sivers function.

COMPASS presented their SSA results with a
polarized deuterium target for Collins and Sivers
moments.\cite{fischer} Both are 
consistent with zero. The difference in comparison
to HERMES  may be explained by a cancellation between
proton and neutron.

\begin{figure}[ctb] 
\begin{center}
\includegraphics[width=.4\linewidth]{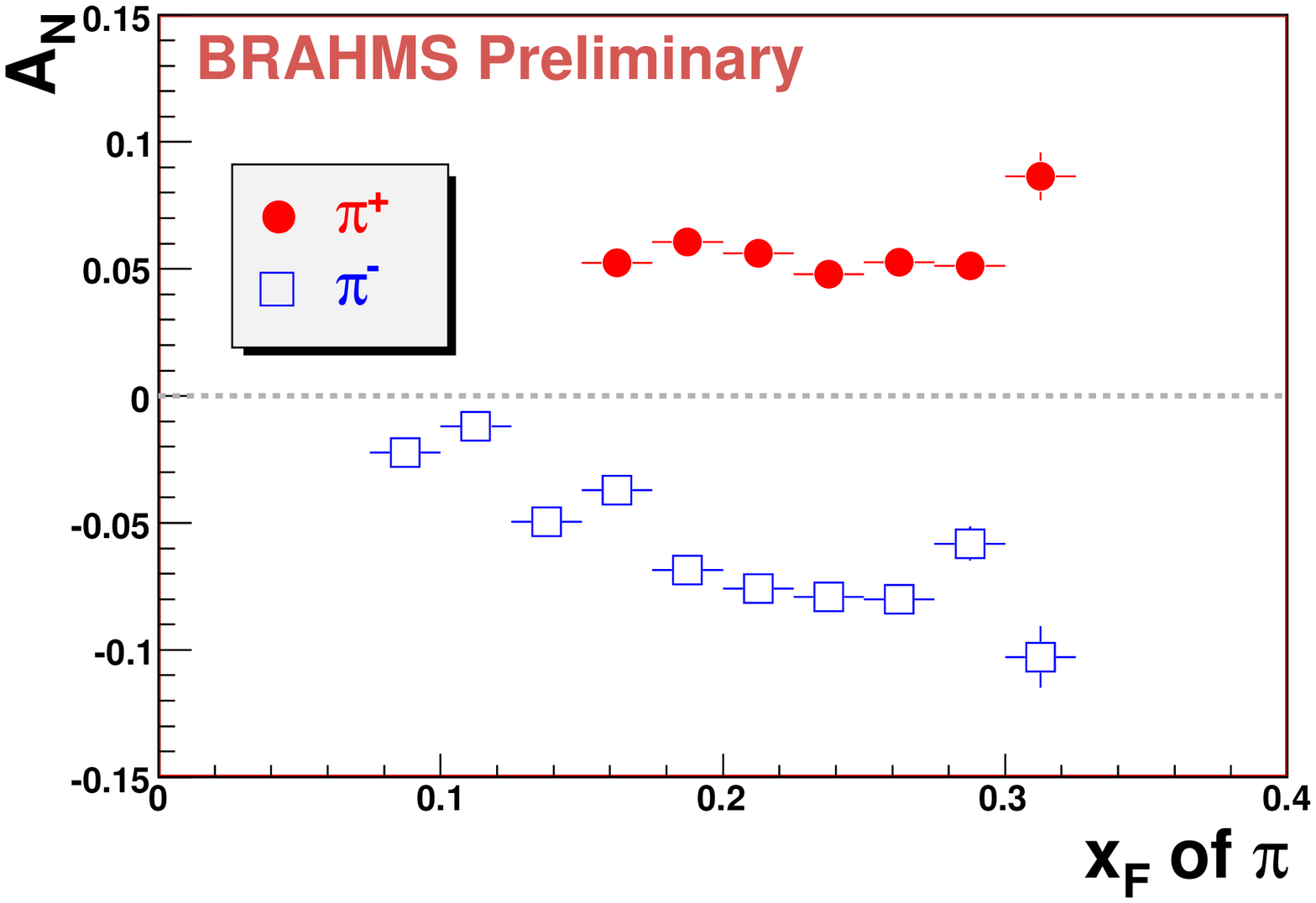}
\includegraphics[width=.4\linewidth]{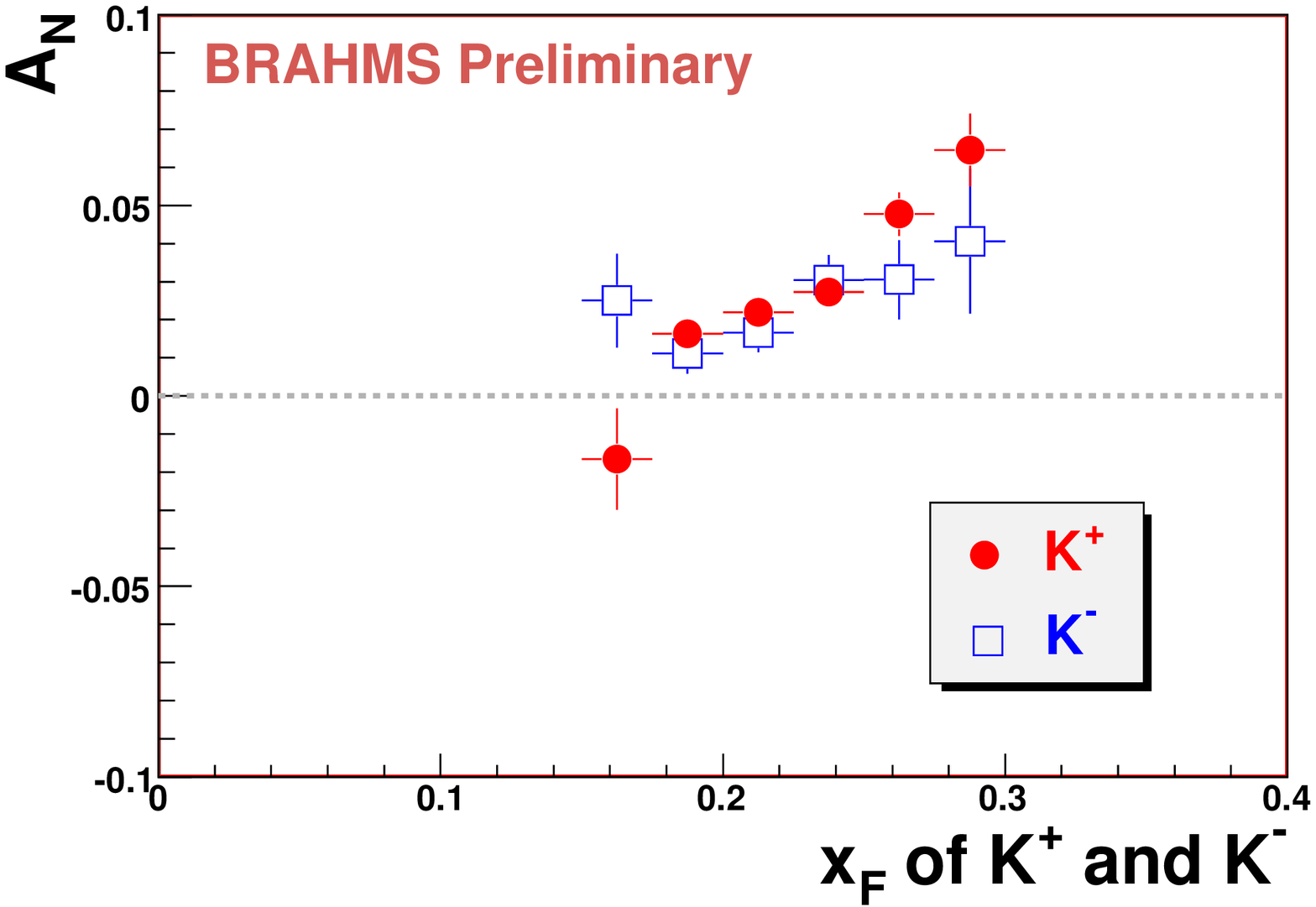}
\end{center}
\caption{Single spin asymmetries  of pions and kaons at BRAHMS.}
\label{fig:BRAHMS}
\vglue-.5cm
\end{figure}

From RHIC, new SSA results, shown in Fig.\ref{fig:BRAHMS}, were presented
by the BRAHMS experiment.\cite{lee}
The SSA of $\pi^+$ ($\pi^-$) displays positive (negative)
asymmetries, of order 5-10\% for $0.1 < x_F < 0.3$.
The SSA of $\pi^+$ for $0.2 < x$ is in agreement with twist-3 calculations.
The SSAs of $K^+$ and $K^-$ are positive and similar to each other,
in disagreement with naive expectation from valence quark
fragmentation.
The SSA for the proton is consistent with zero, while that for
the antiproton is positive. The
cross sections for $\pi^{\pm}$, $K^{\pm}$, proton and
antiproton measured in the same kinematic ranges are  in
agreement with NLO
QCD calculations.

STAR presented $\pi^0$ asymmetries obtained with their
forward and 
backward detectors.\cite{gagliardi}
The asymmetries are positive in the forward region, and consistent with
zero in the backward region.
The forward asymmetries bhave as $1/p_T$, as expected
from perturbative QCD.

PHENIX presented updated asymmetries of charged hadrons at
midrapidity, which are consistent with zero.\cite{tanida}
They also presented neutron asymmetries in the most forward region.
The asymmetry is higher when there is a charged particle activity in the
beam-beam counters.
The SSA is produced via interference of spin flip and spin non-flip
amplitudes.
The one pion exchange model may explain the result, as it aslo explains
the neutron cross sections at ISR.\cite{D'Alesio}

\begin{table}
\tbl{Summary of single spin asymmetry measurements.
 }
{\begin{tabular}{llllll}
\hline
\multicolumn{3}{l}{HERMES} & \multicolumn{3}{l}{COMPASS} \\
proton  & Collins & Sivers & deuteron & Collins & Sivers \\
\hline
$\pi^+$ & $+$     & $+$    & $h^+$    & $0$     & $0$    \\
$\pi^-$ & $-$     & $0$    & $h^-$    & $0$     & $0$    \\
$K^+$   & $0$     & $+$    &          &         &        \\
$K^-$   & $+$     & $0$    &          &         &        \\
\hline
\end{tabular}}
{\begin{tabular}{lllllllll}
\hline
\multicolumn{3}{l}{PHENIX}  & \multicolumn{3}{l}{STAR} & \multicolumn{3}{l}{BRAHMS} \\
\hline
$h^+$   & midrapidity & $0$ &         &          &     & $\pi^+$   & forward  & $+$ \\
$h^-$   & midrapidity & $0$ &         &          &     & $\pi^-$   & forward  & $-$ \\
$\pi^0$ & midrapidity & $0$ & $\pi^0$ & forward  & $+$ & $K^+$     & forward  & $+$ \\
$n$     & zero-degree & $-$ & $\pi^0$ & backward & $0$ & $K^-$     & forward  & $+$ \\
        &             &     &         &          &     & $p$       & forward  & $0$ \\
        &             &     &         &          &     & $\bar{p}$ & forward  & $+$ \\
\hline
\end{tabular}}\label{tblSSA}
\end{table}

The measured SSAs are summarized in Table~\ref{tblSSA}.
A full theoretical understanding of these results is still missing.

Finally, a high-statistics measurement of the single-spin asymmetry of
the proton--proton elastic scattering in the Coulomb Nuclear
Interference (CNI)  region has been
presented by the RHIC polarimeter group.\cite{bravar} The measured
real and imaginary parts of the hadronic spin-flip amplitude, as well
as  the transverse double-spin asymmetries are
consistent with zero, suggesting that 
double spin-flip amplitudes are very small.
New beam-energy
dependence results have been presented for 
the proton--Carbon elastic scattering in the CNI region.

\subsection{Theoretical progress}
\label{transvth}

The measurement of large transverse spin asymmetries is theoretically
challenging, because these asymmetries are negligible in the parton
model:
 a transverse single spin
asymmetry requires a helicity flip, and it is thus
$O\left(\alpha_s\frac{m_q}{\sqrt{Q^2}}\right)$. In perturbative QCD, transverse
asymmetries in the Collins and Sivers processes can  be viewed  in two
distinct ways: from a parton point of view, as a manifestation of
the presence of an intrinsic
dependence of parton distributions on transverse momentum, or from an
operator point of view as the effect of contributions from
twist three quark--gluon
correlation. 

Recent theoretical progress\cite{mulders,yuan}  has led
to a common picture, where these two point of views can be unified,
analogously to what happens for the standard leading--twist collinear
factorization. The basis of the unification is
a relation 
between the transverse--momentum dependent quark distribution, and the
relevant twist three operator. Specifically a transverse cross section
difference (e.g. in SIDIS or Drell-Yan)
can be schematically factorized as 
\begin{equation}
\label{trfact}
\Delta d \sigma\sim \epsilon_{\alpha\beta} s_\perp^\alpha
p_\perp^\beta
\int\frac{dx}{x}\int\frac{dz}{z}q(z) T_F(x,x-xg),
\end{equation}
where $q(z)$ is a conventional (collinear) 
quark distribution and $T_F(x_1,x_2)$ is a
twist--three quark-gluon correlation. It can then be
shown\cite{mulders} that the quark--gluon correlation is related by
$T_F(x,x)=\int d^2k_\perp | \vec k_\perp|^2 q_T(\vec k_\perp,x)$
to the transverse--momentum dependent quark distribution $q_T(\vec k_\perp,x)$, defined in
terms of a suitable nucleon matrix element of a quark-quark bilinear
connected by a gauge link.

One can show\cite{yuan} that when $k_\perp<\!\!<Q$
the single--spin asymmetries can be factorized in terms of $q_T(\vec k_\perp,x)$,
convoluted with a transverse fragmentation function (Sivers function),
and a perturbatively
computable factor related to soft gluon radiation. The
$k_\perp$ dependence of  $q_T(\vec k_\perp,x)$ can then be
computed perturbatively. Substituting the result of the latter
computation
in the former factorized
expression,  the expression
eq.~(\ref{trfact}) for the cross section is reobtained, thus
showing the equivalence of the two approaches. Furthermore, it can be
shown that the transversity structure is universal:\cite{mulders}  for
instance, the Sivers functions for 
Drell-Yan and  SIDIS are both given in terms of a single process-independent
function, determined by partonic matrix elements. In fact, the
Boer--Mulders $h_1^\perp$ and Sivers $f_{1T}^\perp$ functions can all be
expressed in terms of $2n_f+1$ universal quark and gluon matrix
elements.
These results impose powerful constraints on phenomenological studies of
single spin asymmetries, and they can guide the construction of
phenomenological models\cite{gamberg} for the Sivers function.

\section{Stretching the boundaries}
\label{stretch}
The widening of the scope of spin physics has led to an extension  to the
polarized case of lines of experimentation and theoretical analysis
which hitherto had been explored only at the unpolarized level.

\subsection{Fragmentation}
\label{frag}

The BELLE experiment has measured a significant non-zero asymmetry in the
double ratio of unlike-sign pion pairs to like-sign pion pairs (UL/L)
produced from $e^+ e^- \rightarrow q \bar{q}$ reactions in the off-resonance
region.\cite{seidl}
The asymmetry is sensitive to the Collins fragmentation function, but it is
not very sensitive to the favored to disfavored Collins function ratio.
A 
new double ratio of unlike-sign pion pairs to charged pion
pairs (UL/C), which is sensitive to the favored + unfavored Collins
function, has been measured to be about  half of the UL/L asymmetry.

The COMPASS experiment has measured both longitudinal and transverse
polarization transfers of $\Lambda$ and $\bar{\Lambda}$
production.\cite{grube}
By averaging over the target polarization, they determine 
the polarized fragmentation functions, $\Delta D_{\Lambda/q}(z_h)$.
The longitudinal polarization transfer provides a test of $q \bar{q}$ symmetry of
the polarized strange sea in the nucleon.
Results shows similar longitudinal polarization for $\Lambda$ and
$\bar{\Lambda}$ in spite of different production mechanism.
The transverse polarization transfer gives information on initial transverse
quark polarization $\Delta q_T(x)$ in the nucleon.
The result shows a slight tendency towards negative polarization transfer,
and   a small positive spontaneous transverse
polarization of $\Lambda$ and unpolarized $\bar{\Lambda}$.


\subsection{Structure functions at low $Q^2$}
\label{lowq}

Structure function measurements have been recently  
extended to the low-$Q^2$ region. At low $Q^2$ 
and low $x$, $g_1$   can be compared to predictions from  Regge theory, VMD
and low--energy models.
The COMPASS experiment presented a high precision $A_1^d$ measurement at
$Q^2 <$ 1 GeV$^2$  $0.00005 < x < 0.02$.\cite{fischer} 
The measured $A_1^d$ and $g_1^d$ are compatible with zero.

The JLab experiments are measuring $g_1$ in the nucleon resonance
region, thereby investigating parton-hadron duality, both global
(integrated in $x$) and local in $x$.
The CLAS experiment in Hall B has investigated the duality property of
$g_1^p$ and $g_1^d$ at low $Q^2$.\cite{dharmawardane}
Quark-hadron duality seems to be supported at the global level, and 
and also locally in some of the resonance
regions.
The E01-012 experiment at Hall A is measuring $g_1$ and $g_2$ on the $^3$He
target,\cite{liyanage} and the RSS experiment at Hall C is measuring them
on proton and deuteron targets.\cite{tajima}

\subsection{Resummation}
\label{resum}

Current and future
polarized experiments will involve processes and kinematical
regions where fixed order computations are not sufficient, and this has
stimulated the extension to the polarized case of  resummation
techniques: specifically,
resummation of an inclusive process close to its kinematic threshold, such as
Drell-Yan when $Q^2\to s$, and resummation of the $p_T$ distribution
at small $p_T$.

Threshold resummation for the polarized Drell-Yan process is important
for future experiments at J-PARC and GSI.
Threshold resummation up to the
next-to-leading logarithmic (NLL) level for the
transverse  Drell-Yan  spin asymmetries have been performed in
Ref.\cite{yokoya}, both inclusive  and differential in
rapidity; and up to NNLL in the unpolarized case. The resummed
$K$-factors are large, leading to an increase of the cross section by a
large factor for invariant masses above $\sim4$~GeV$^2$, 
but essentially spin independent, so that the asymmetry
is only moderately affected.

The resummation of the $q_T$ distribution for the transversely
polarized Drell-Yan process is necessary even at RHIC energies,
because the unresummed cross section diverges as $q_T\to 0$. A
determination of the resummed cross section up to the NLL level~\cite{kawamura}
shows that unresummed result are in fact unreliable even
for intermediate values of $q_T\sim Q/4$, where the cross--section
difference is peaked: the resummation gives the dominant
contribution, and unresummed results are reproduced only for large
$q_T\sim Q$, where the cross section is very small. 

The resummation of the $q_T$ distribution has also been performed~\cite{koike} for
the $q_T$ spectrum of single--inclusive hadron production in
DIS, in the unpolarized case, for
longitudinally polarized electron and proton, for longitudinally
polarized incoming electron and outgoing hadron, and both for
longitudinally and transversely polarized incoming proton and outgoing
hadron. The case of longitudinal $ep$ polarization is relevant for the
COMPASS and HERMES experiments discussed in Sect.~\ref{spinpuz}. In
this case, one finds that, for the kinematics of these experiments,
the impact of resummation effects is again rather large, but it
largely cancels in the asymmetry.

A difficulty in the determination of resummed results is due to ambiguities caused by the
fact that at the resummed level the strong coupling hits the Landau
pole. These ambiguities are moderate in
threshold resummation, but become more important in $q_T$ resummation.
For the case of SIDIS, the
ambiguity can be as large as the whole resummation, which suggests
that a purely perturbative treatment of the process is not really
possible, and further undermines its usefulness for determinations of
the polarized hadron structure.
\vglue-.5cm
\section{Outlook}

Considerable progress is expected in the near future thanks to the
completion of the COMPASS and especially RHIC experimental programs.
On top of the forthcoming determinations of $\Delta G$ discussed 
in Sect.~\ref{spinpuz},
RHIC experiments will determine the
flavor decomposition of quark and antiquark
polarization through $W$ production
at $\sqrt{s} = 500$~GeV.\cite{liu,simon}
Future experiments at J-PARC and GSI will further explore longitudinal and
transverse polarized distributions of quarks and antiquarks through
the  Drell-Yan
process.\cite{lenisa}

\end{document}